\documentclass{aa}
\usepackage{epsfig}
\usepackage{psfig}
\usepackage{graphicx}
\usepackage{txfonts}

\usepackage{lscape}

\newcommand{\PSR}{PSR~B1929$+$10}
\titlerunning{Resolving the bow-shock nebula around \PSR}

\begin{document}
\title{Resolving the bow-shock nebula around the old pulsar \PSR\ with 
multi-epoch Chandra observations}

\author{C. Y. Hui \and W. Becker}   
\date{Received 06 July 2007 / Accepted 28 April 2008}
\institute{Max-Planck Institut f\"ur Extraterrestrische Physik, 
          Giessenbachstrasse 1, 85741 Garching bei M\"unchen, Germany}

\abstract{We have studied the nearby old pulsar \PSR\ and its surrounding
 interstellar medium utilizing the sub-arcsecond angular resolution of the Chandra X-ray Observatory.
 The Chandra data are found to be fully consistent with the results obtained 
 from deep XMM-Newton observations as far as the pulsar is concerned.  We confirm  
 the non-thermal emission nature of the pulsar's X-radiation. In 
 addition to the X-ray trail already seen in previous observations by the ROSAT 
 and XMM-Newton X-ray observatories, we discovered an arc-like nebula surrounding the pulsar.  
 We interpret the feature as a bow-shock nebula and discuss its energetics  
 in the context of standard shock theory.

}

\maketitle

\section{Introduction}
 In 1990s, a series of powerful X-ray observatories
 were launched to space. Among many great results on various aspects in
 astrophysics, these observatories allowed a first detailed study of the X-ray emission 
 properties of rotation-powered pulsars as a class (cf.,~Becker \& Tr\"umper 
 1997; Becker \& Pavlov 2001, for a review). Most old\footnote{In standards
 of high energy astronomy rotation-powered pulsars are called young, middle 
 aged, and old if their spin-down age is of the order of few times $10^3-10^4$ 
 yrs, $10^5-10^6$ yrs, and $\ge 10^6$ yrs, respectively. This classification 
 is diffuse, though, with a smooth transition in between the different groups.} 
 radio pulsars, though, were still too faint for a detailed study by these 
 satellites (cf.,~Sun et al.~1993; Manning \& Willmore 1994; Becker \& Tr\"umper 
 1997; Saito 1998).
 Thanks to the much improved sensitivity of the XMM-Newton and Chandra 
 observatories, a more detailed study of old pulsars became possible.
 Observations of PSR B0950+08; B0823+26; J2043+2740 (Becker at 
 al.~2004; Zavlin \& Pavlov 2004); B0628-28 (Becker et al.~2005); 
 B0943+10 (Zhang, Sanwal \& Pavlov 2005); B1133+16 (Kargaltsev, Pavlov \& 
 Garmire 2006); B1929+10 (Becker et al.~2006); and B2224+65 (Hui \& Becker 2007a) 
 for the first time have allowed us to study the emission properties of 
 old pulsars as a class. Surprisingly, the X-ray emission from old pulsars 
 seems to be dominated by non-thermal radiation processes. Thermal components 
 (e.g.,~to account for the emission from hot polar caps) are not required to 
 model the X-ray spectra of these pulsars, resulting in mostly upper limits 
 for any thermal radiation components. Further support for an emission scenario 
 dominated by non-thermal radiation is found by the observed temporal emission 
 properties. Pulse profiles, if detected with sufficient photon statistics,  
 show multiple components and narrow features. This is indicative of strongly 
 beamed emission, which further invalidates the heated polar cap scenario as 
 the main source of X-ray emission in old pulsars. The pulsed fractions 
 in old pulsars are in the range of $\sim 30-50\%$.

 In addition to the pulsar emission, which originates within the co-rotating 
 magnetosphere, extended trail like X-ray emission was observed from \PSR\ 
 (Becker et al.~2006) and PSR B2224+65 (Hui \& Becker 2007a) on a scale of 
 several arc-minutes. In the case of \PSR\ the X-ray emission in the trail is 
 interpreted as synchrotron emission produced in the shock between the pulsar 
 wind and the surrounding medium (see the discussion in Becker et al.~2006 for 
 details). Owing to the moderate  (15" half energy width) spatial resolution 
 of XMM-Newton, details of the nebular emission associated with \PSR\ might 
 have been remained unresolved. With a ten times improved angular resolution 
 Chandra data can  thus be essential to further constrain the properties of 
 the nebula very near to the pulsar. 

 In this paper, we present a detailed analysis of multi-epoch Chandra 
 observations of the field around \PSR. 
 Based on its X-ray emission properties, this source is considered 
 prototypical of an old pulsar (Becker et al.~2006). With a pulse period 
 of $P=226.5$~ms and a period derivative of $\dot{P}=1.16\times 10^{-15}$, 
 its characteristic age is determined to be $\sim 3\times 10^6$ years. These 
 spin parameters imply a spin-down luminosity of $\dot{E} = 3.9 \times 10^{33}\, 
 \mbox{erg s}^{-1}$ and a magnetic field at the neutron star magnetic poles of\,
 $B_\perp \sim 5 \times 10^{11}$~G.  With a radio dispersion measure of $\,3.178\,
 \mbox{pc cm}^{-3}$, the NE2001 Galactic free electron density model of 
 Cordes \& Lazio (2002) predicts a distance of 170 pc.  However, the  recent 
 astrometric measurements by Chatterjee et al.~(2004) yielded a precise proper 
 motion and parallax determination that translates into an accurate distance 
 measurement of $d=361^{+10}_{-8}$ pc and a proper motion of
 $V_\perp=177^{+4}_{-5}$ km s$^{-1}$. The ephemerides of \PSR, which we made 
 use of in this paper, are listed in Table 1.

 The paper is organized as follows: in Sect. 2 we describe the observations and 
 data analysis; and in Sect. 3 we summarize the results and provide a discussion 
 in the context of a standard shock model for the nebular emission. 

\section{Observations and data analysis}

\begin{figure*}
\psfig{figure=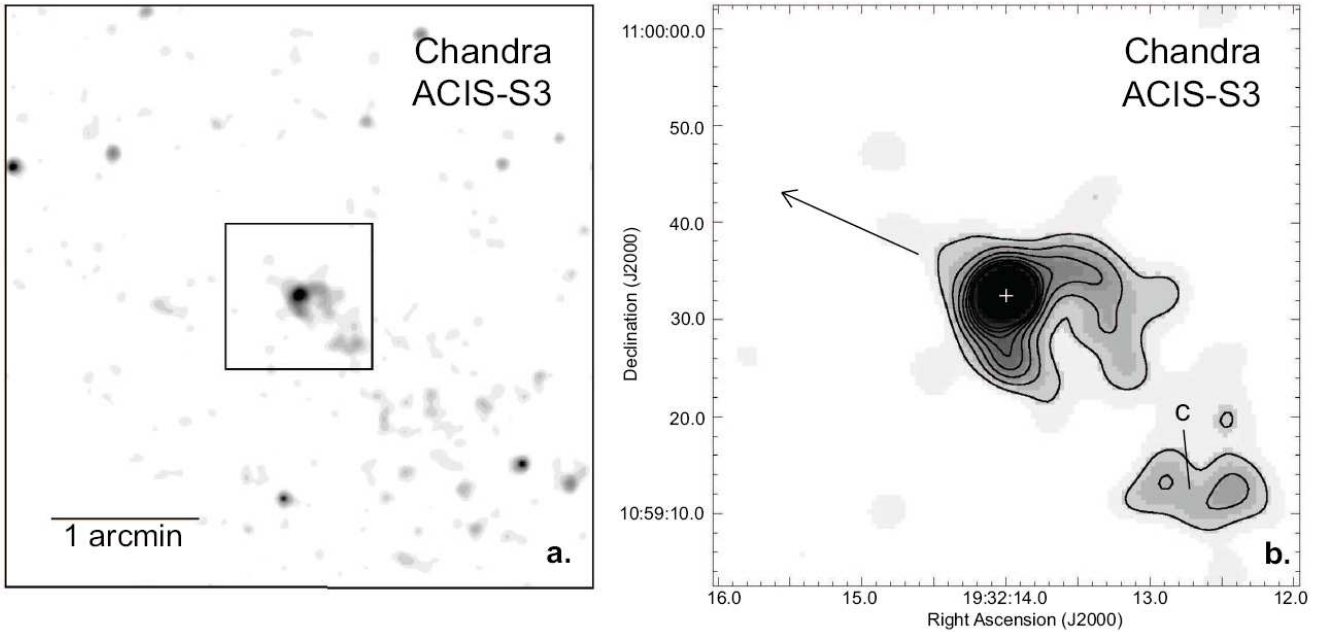,width=18cm,clip=}
\caption{{\bf (a)~} $4\times4$ arcmin image centered on \PSR\ 
in the energy band of $0.5-8$ keV. Top is north
and left is east. The image is created by merging both Chandra observations.  
It is binned with a factor of 0.5 arcsec and adaptively
smoothed with a kernel of $\sigma < 2$ arcsec. 
{\bf (b)~} The close-up view of the central $1\times 1$ arcmin region as 
illustrated in {\bf a}. The proper motion corrected pulsar position is indicated by 
the white cross. 
An arc-like feature associated with the pulsar is observed. The proper motion 
direction of the pulsar (Chatterjee et al. 2004) is illustrated by the arrow. 
Contour lines at the levels of 
$(6.1-26)\times10^{-6}$ counts arcsec$^{-2}$ s$^{-1}$ are overlaid.}
\end{figure*}

\begin{table}
\caption{Ephemerides of \PSR\ $^{a}$}
\begin{center}
\begin{tabular}{lc}
\hline\hline
       &   \\
\hline
Right Ascension (J2000)  & $19^{\rm h} 32^{\rm m} 13.983^{\rm s}\pm
0.002^{\rm s}$ \\
Declination (J2000)          & $+10^\circ\; 59'\; 32.41"\pm 0.07"$  \\
First date for valid parameters (MJD) & 52929 \\
Last date for valid parameters (MJD) & 53159 \\
Pulsar rotation period (s)        & 0.2265182954 \\
Period derivative $\dot{P}$ ($10^{-15}$ s s$^{-1}$) & 1.164739 \\
Characteristic age ($10^{6}$ yrs)          & 3.09 \\
Surface dipole magnetic field ($10^{12}$ G) & 0.5129 \\
Dispersion Measure (pc cm$^{-3}$)  & 3.178 \\
Distance (pc)   &  $361^{+10}_{-8}$ \\
Spin-down Luminosity ($10^{33}$) ergs s$^{-1}$ & 3.89 \\
\hline
\end{tabular}
\end{center}
$^{a}$Adopted from Becker et al. (2006)\\
\end{table}

 Data analysis is restricted to the energy range $0.5-8.0$ keV. All energy 
 fluxes, however, are computed for the $0.5-10$ keV band for better comparison 
 with the results based on XMM-Newton data (Becker et al.~2006).

 The object \PSR\ was observed with Chandra in 2005 December 04 (Obs ID: 6657) 
 and 2006 May 28 (Obs ID: 7230) with the Advanced CCD Imaging Spectrometer 
 (ACIS). In both observations, \PSR\ was located on the back-illuminated (BI) 
 ACIS-S3 chip with an off-axis angle of $\sim 0.1$ arcmin. We used standard processed 
 level-2 data. The effective exposures are $\sim 21$ ks and $\sim 25$ ks 
 for the observations in 2005 December and 2006 May, respectively. 

 \subsection{Spatial analysis}

 With the objective obtaining better statistics for the analysis, we combined 
 both datasets to produce better images. We carefully 
 checked and corrected the aspect offsets for each observation prior to the merging. 

 The combined X-ray images of the $4\times4$ arcmin field centered on \PSR\ as well
 as a close-up of the central $1\times1$ arcmin regions are shown in Fig.~1.  
 A compact nebula, which has an arc-like morphology somewhat resembling a bow-shock,   
 is clearly detected around \PSR. X-ray contours were calculated at the levels of 
 $(6-26)\times10^{-6}$ counts arcsec$^{-2}$ s$^{-1}$ and overlaid on the image in 
 Fig. 1b.  Assuming this arc-like compact feature is a bow-shock pulsar wind 
 nebula, it is interesting to estimate the stand-off angle $\theta_{s}$ which is 
 a crucial parameter in discussing the shock physics (see also Sect. 3). By fitting 
 a 2-D Gaussian model to the raw image of the field around \PSR, we obtained 
 a full width at half maximum (FWHM) of $1.13\pm0.02$ arcsec ($1\sigma$ 
 error) . We take this as an estimate for the lower bound of $\theta_{s}$.

 To further examine the extent of the compact nebula, we computed the brightness 
 profile in the energy band $0.5-8$ keV  from the raw image of bin size of 0.5 
 arcsec. We estimated the counts in sixteen consecutive boxes of dimensions $1\times15$ 
 arcsec, oriented along the nebula. The vignetting corrected  brightness profile 
 and the orientation of the 16 spatial bins are shown in Fig.~2. 
 The arc-like nebula extends up to $\sim10$ arcsec from the pulsar before 
 fading into the background. Because of the wings of the ACIS-S 
 point spread function, the pulsar is expected to contribute to its surrounding 
 nebula. To estimate this, we compared the observed data with the
 ACIS-S instrumental point spread function. To do so, we simulated a point source observation 
 using MARX (ver 4.2.1). The adopted inputs were the pulsar spectrum and the same 
 exposure, roll, and off-axis angle as in the actual data.
 We then computed the brightness profile of the simulated data in the same way as
 aforementioned. The estimated point source profile is shown in Fig.~2
 as a dotted line. The brightness profile of the arc-like nebula computed from the
 actual observation is significantly different from the expected distribution
 of a point source. On the other hand, there is no significant extended emission
 detected ahead of \PSR.

\begin{figure}[h]
\psfig{figure=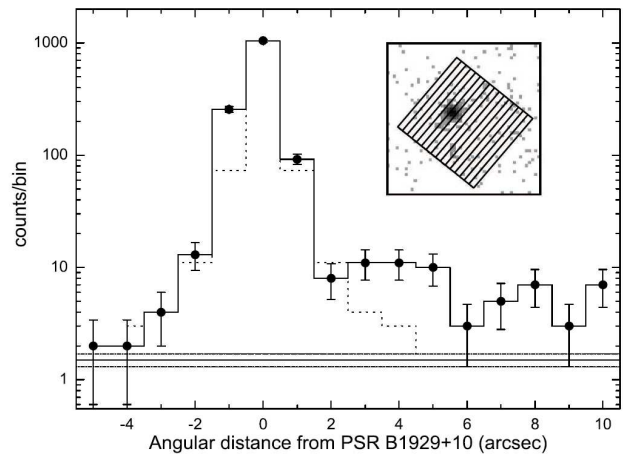,width=8.5cm,clip=}
\caption{Vignetting corrected brightness profile of the field around \PSR.
 The energy range is 0.5$-$8 keV. The profile expected for a point source is
 indicated by a dotted line. The average background level and its $1\sigma$
 deviation are indicated by horizontal lines, which were calculated by sampling 
 the source-free regions around the pulsar within a $1\times1$ arcmin 
 field-of-view. The inset shows the sixteen
 bins used in computing the profile. Each bin is $1"\times15"$.}
\end{figure}

 In addition to the arc-like nebula near the pulsar, we observed a clumpy structure (which is 
 labeled as  C in Fig.~1b). There are only $\sim 24$ net counts from 
 this clump, which does not allow for a more detailed analysis. The signal-to-noise 
 ratio of this clump is estimated to be $\sim 2$, and thus is consistent with 
 background fluctuations. 

 In Fig.~1a, no prominent structure resembling the X-ray trail 
 seen by ROSAT and XMM-Newton opposite to the pulsar's proper motion
 direction can be identified (cf.~Fig.~2 in Becker et al.~2006). 
 We have examined the Chandra image further by smoothing the 
 raw data with a kernel of $\sigma <6$ arcsec, which is comparable with the FWHM 
 of XMM-Newton's PSF. The smoothed image is displayed in Fig.~3. Comparing 
 the image with the contours calculated from XMM-Newton MOS1/2 data, a faint 
 trail like feature is noticed. However, the background contribution of this 
 feature is estimated to be $\sim 60\%$. The unabsorbed X-ray flux of a 1
 arcmin circular trail region near the pulsar, as detected by XMM-Newton, is 
 $f_X=5.3\times 10^{-14}\,\mbox{erg/s/cm}^2$ within the 0.5--10 keV band 
 (Becker et al.~2006). The low significance of the trail in the two Chandra 
 observations, thus, is in agreement with Chandra's lower sensitivity. We,  
 therefore, will not further consider the X-ray trail emission in this paper, and
 alert the interested reader to the XMM-Newton results reported by 
 Becker et al.~(2006), for a detailed discussion of its emission properties. 

\begin{figure}
\psfig{figure=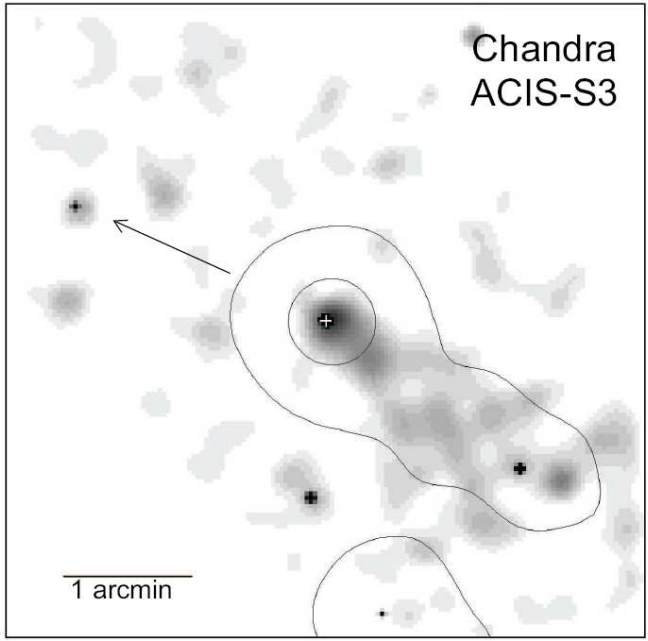,width=9cm,clip=}
\caption{The $5\times5$ arcmin region around \PSR\
in the energy band of $0.5-8$ keV. The image is created by merging both 
Chandra observations. It has a binning factor of 2 arcsec and is adaptively
smoothed with a kernel of $\sigma < 6$ arcsec. Top is north and left is east. 
The contours from the XMM-Newton's MOS1/2 $0.2-10$ keV band image at the levels  
of 0.09 and 0.44 counts arcsec$^{-2}$ are overlaid. The proper motion corrected 
pulsar position is indicated by the white cross. The proper motion direction 
of the pulsar is indicated by the arrow in the image. }
\end{figure}

 \subsection{Spectral analysis}

 Although the spectrum from \PSR\ has already been tightly constrained by the 
 XMM-Newton data (Becker et al.~2006) it is important for us to re-examine 
 its spectral properties with Chandra since it resolves the pulsar 
 emission from the compact surrounding nebular component better.

 We extracted the pulsar spectrum in both data sets from a circle of 2 arcsec 
 radius (encircled energy $\sim$95\%), centered on the pulsar, and fitted them 
 simultaneously. The background spectrum was extracted from a source-free region 
 within a 10 arcsec radius centered at R.A.=$19^{\rm h}32^{\rm m}14.107^{\rm s}$, 
 Dec.=$10^{\circ}59'51.57"$ (J2000).  After background subtraction, $593\pm24$ and 
 $682\pm26$ net counts were available for the spectral analysis. These values 
 imply the net counting rates of $(2.84\pm0.11)\times 10^{-2}$ cts/s and 
 $(2.77\pm0.11)\times 10^{-2}$ cts/s for the observations in 2005 December and 2006 
 May, respectively. 

 We computed the response files with the tools MKRMF and MKARF in CIAO 3.4. 
 Utilizing the most updated calibration data, CALDB 3.4.1, the generated response 
 files corrected the degradation of the quantum efficiency in the ACIS CCD, accordingly. 
 Each spectrum was dynamically binned so as to have at least 30 counts per bin. 
 To better constrain the spectral properties, we fitted simultaneously the spectra obtained from both 
 observations. We performed all the spectral fittings in $0.5-8$ keV with 
 XSPEC 11.3.2. The parameters of all fitted model spectra are summarized in Table 2. 
 All the quoted errors are $1-\sigma$, and were computed for 1 parameter of interest.

 Among the single component models tested, we found that a power-law model fits 
 the data best ($\chi^{2}_{\nu}=0.89$ for 36 D.O.F.). This model yields a column 
 density of $N_{H}=2.40^{+0.36}_{-0.34}\times10^{21}$ cm$^{-2}$, a photon index of\/ 
 $\Gamma=2.91^{+0.16}_{-0.13}$, and a normalization at 1 keV of $8.12^{+1.04}_{-0.90}
 \times 10^{-5}$ photons keV$^{-1}$ cm$^{-2}$ s$^{-1}$. These best-fit values are 
 fully consistent with those obtained by XMM-Newton for power-law fits (see Table 3 
 in Becker et al.~2006). The best-fit power-law spectrum and residuals are shown in 
 Fig.~4. The column density inferred from both Chandra and XMM-Newton spectra is 
 relatively high in comparison with the mean value obtained from the Extreme UltraViolet 
 Explorer (EUVE) measurements of the stars in the neighborhood of the pulsar 
 (Slowikowska et al. 2005). However, as discussed by Slowikowska et al.~2005,  
 the interstellar medium is very patchy in the region around \PSR\ (cf.,~Table~3 in 
 Slowikowska et al. 2005 for more details). Merely on the basis of the comparison 
 with the mean value,  the $N_{H}$ we report and in Becker et al.~(2006) 
 cannot be invalidated. A more detailed mapping is thus needed to better constrain 
 the $N_{H}$ toward the pulsar.

 We have also computed the error contours to demonstrate the relative parameter 
 dependences of the photon index vs.~the hydrogen column density and plotted this
 in Fig.~5.  The unabsorbed flux deduced for the best fit power-law model 
 parameters is $f_{X}=2.5\times10^{-13}$ ergs s$^{-1}$ cm$^{-2}$ within $0.5-10$ keV. 
 At a distance of 361 pc it implies a luminosity of $L_{X}=3.9\times10^{30}$ ergs s$^{-1}$. 

 As for the XMM-Newton data it is obvious that the single power-law model already 
 describes the observed pulsar spectrum very well. Hence, the justification of 
 including an additional thermal component is absent. In the Chandra data, we 
 found that fitting with a power-law plus blackbody model does not yield a reasonable 
 solution if we allow the blackbody radius and the temperature to be free parameters. 
 It resulted in a blackbody radius of $R_{\rm bb}=25.81^{+18.34}_{-25.81}$ m. 
 This is much smaller than the canonical size of a polar cap, 
 $r_{\rm pc}=R(2\pi R/cP)^{1/2}\sim 300$~m, which suggests the contribution from 
 the additional blackbody component is insignificant. We quantified the statistical 
 significance for adding this extra component to the power-law model with the 
 $F-$test which suggests that inclusion of this thermal components is only required 
 at a confidence level of $26\%$. 

 We have also examined the possible contributions from a polar cap and the neutron 
 star surface by fixing the blackbody radii at $R=300$ m and $R=10$ km, respectively. 
 These fits yield temperatures of $T\sim8.7\times 10^{5}$ K and $T\sim 5.6\times 10^{5}$ K.  
 The $F-$test suggests that adding these thermal components to the power-law model 
 is only significant at a confidence level $\la 62\%$. 
 Due to the better photon-statistics this number was even 
 smaller in the XMM-Newton data (Becker et al.~2006). This low significance 
 is also reflected by the relative contribution of the thermal component in 
 the total energy flux observed by Chandra. We estimated the $1\sigma$ 
 upper limits for the polar cap temperature and its flux contribution 
 by adding a blackbody component to the power-law model. 
 Fixing the blackbody radius at a polar cap size of $R=300$ m, a $1\sigma$ 
 upper limit of $T<1.2\times10^{6}$ K was computed by assuming contribution from one polar cap 
 only. This implies that the 1$\sigma$ limit contribution from the polar cap in the total energy flux is 
 $\sim 12\%$. 

 For the spectral model consisting of two blackbody components, we found that the 
 best-fit model parameters also agree well with those inferred from the XMM-Newton 
 spectra (Becker et al.~2006). Despite the acceptable value of the goodness-of-fit, 
 the inferred blackbody radii are too small to be considered as a reasonable description. 
 Moreover, the pulsar spectrum obtained by XMM-Newton and fitted together with the
 spectrum obtained by the ROSAT PSPC have already shown that this model cannot 
 describe the data beyond $\sim 5$ keV (see Becker et al.~2006). 

\begin{table*}
\caption{Spectral parameters inferred from fitting the Chandra ACIS-S data
obtained from \PSR\ and its associated extended feature.}
\begin{center}
\begin{tabular}{lcccccc}
\hline\hline
Model $^{a}$ & $\chi^{2}_{\nu}$ & D.O.F. & $N_{H}$ & $\Gamma^{b}$ / $kT$ & Normalization at 1 keV$^{c}$ & Radius $^{d}$\\
      &                  &        & $10^{21}$ cm$^{-2}$ &   & photons keV$^{-1}$ cm$^{-2}$ s$^{-1}$  & m \\
 \hline
 \multicolumn{7}{c}{\PSR}\\
 \hline\\
 PL    & 0.89 & 36 & $2.40^{+0.36}_{-0.34}$  &  $2.91^{+0.16}_{-0.13}$   & $8.12^{+1.04}_{-0.90}\times10^{-5}$   & - \\
 \\
 BB    & 2.00  & 36  & $<0.04$  & $0.36^{+0.01}_{-0.01}$  & - & $29^{+2}_{-2}$ \\
 \\
 PL+BB   & 0.93  & 34 & $1.40^{+1.09}_{-l.40}$  &  $2.48^{+0.52}_{-1.09}$/$0.29^{+0.05}_{-0.29}$  & $4.60^{+6.29}_{-3.42}\times10^{-5}$ & $25.81^{+18.34}_{-25.81}$ \\
 \\
 PL+BB   & 0.92  & 35 & $2.44^{+0.60}_{-0.39}$  &  $2.91^{+0.16}_{-0.16}$/$0.07^{+0.03}_{-0.07}$  & $8.17^{+1.09}_{-0.94}\times10^{-5}$ & $300$ \\
 \\
 PL+BB   & 0.90  & 35 & $2.72^{+1.06}_{-0.55}$  &  $2.97^{+0.27}_{-0.17}$/$0.05^{+0.01}_{-0.05}$  & $8.67^{+2.25}_{-1.22}\times10^{-5}$ & $10000$ \\
 \\
 BB+BB   & 0.91 & 34  & $<0.70$  &  $0.71^{+0.21}_{-0.13}$/$0.27^{+0.03}_{-0.05}$  & - & $5.37^{+3.43}_{-2.30}$/$42.32^{+22.74}_{-4.47}$  \\
 \\
      \hline
      \multicolumn{7}{c}{Arc-like feature}\\
      \hline\\
      PL    & 0.88  & 6  & $2.40$  &  $2.00^{+0.32}_{-0.30}$   & $3.63^{+0.78}_{-0.74}\times10^{-6}$ & -  \\
      \\
\hline\hline
 \end{tabular}
 \end{center}
 $^{a}$   {\footnotesize PL = power-law; BB = blackbody}\\
 $^{b}$   {\footnotesize The entry in this column depends on the model in interest. It is 
 the temperature $kT$ in keV or the photon index $\Gamma$}\\
 $^{c}$   {\footnotesize The normalization constant for the power-law model.}\\
 $^{d}$   {\footnotesize The radius of the blackbody emitting area is 
 calculate for an assumed pulsar distance of 361 pc.}\\
  \end{table*}

 \begin{figure}
 \centerline{\psfig{figure=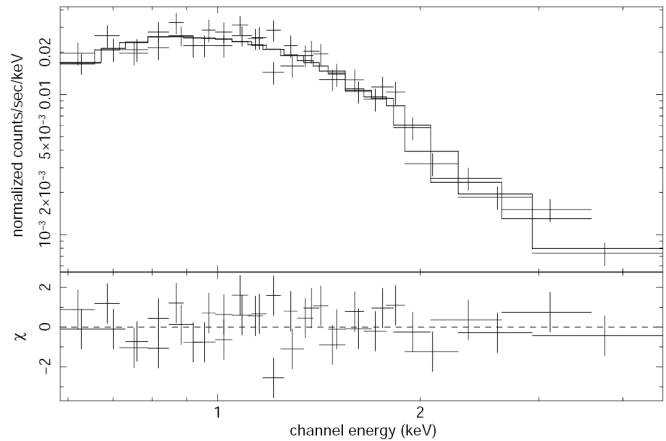,width=9cm,clip=}}
 \caption{Energy spectra of \PSR\ as observed with the Chandra
 ACIS-S3 detector on 04 Dec 2005 and 28 May 2006 and simultaneously 
 fitted to an absorbed power-law model
 ({\it upper panel}) and contribution to the $\chi^{2}$ fit statistic
 ({\it lower panel}).}
 \end{figure}

 \begin{figure}
 \centerline{\psfig{figure=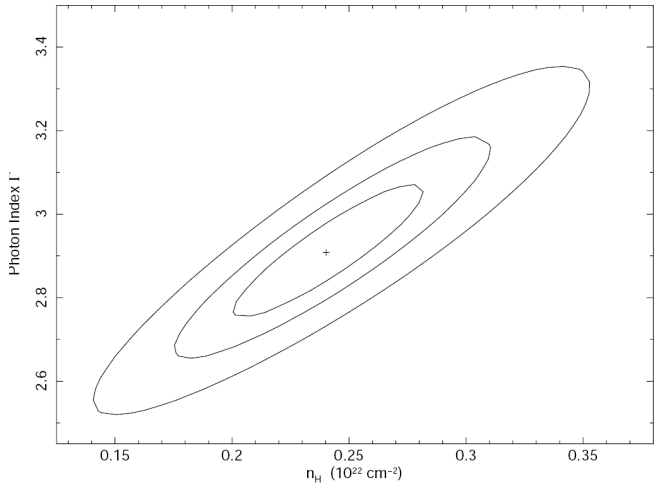,width=9cm,clip=}}
 \caption{$1\sigma$, $2\sigma$ and $3\sigma$ confidence contours calculated 
 for 1 parameter of interest for the power-law model fitted to the 
 spectrum of \PSR.}
 \end{figure}

 \begin{figure}
 \centerline{\psfig{figure=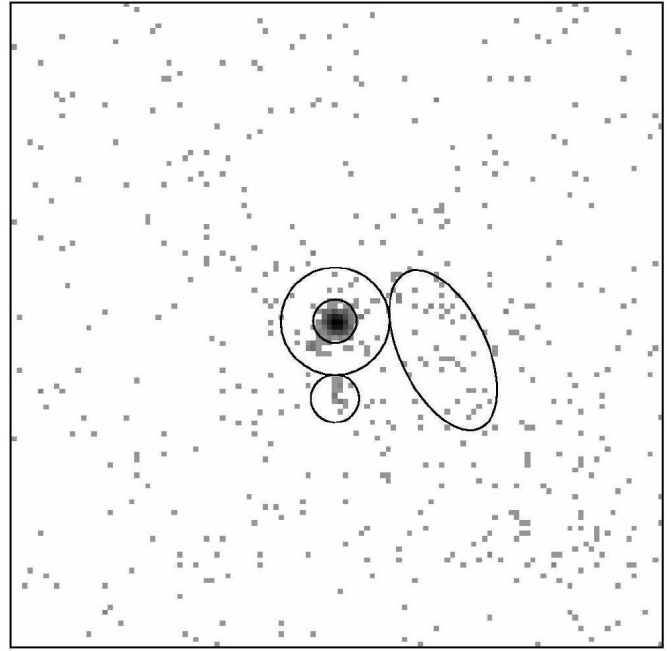,width=9cm,clip=}}
 \caption{$1\times1$ arcmin raw image ($0.5-8$ keV) centered on \PSR. 
  Data from both Chandra observations were used. The extraction regions 
  used for the spectral analysis of the arc-like feature are illustrated.}
 \end{figure}

 Since the arc-like nebula is resolved by Chandra for the first time, 
 it is instructive to examine its energy spectrum, despite the poor photon 
 statistics. We extracted the nebular spectra  from the regions 
 in accordance with its observed morphology. For the sake of consistency, 
 we adopted the same extraction regions in both datasets, which are 
 illustrated in Fig.~6. It consists of an annular region centered on 
 the pulsar position with inner and outer radii of 2 arcsec and 5 arcsec, 
 a circular region with a radius of 2.2 arcsec, as well as a ellipse 
 of 4 arcsec$\times$8 arcsec. 

 The background spectra were extracted from each dataset within a nearby 
 source-free region of a 10 arcsec radius, centered at RA.=$19^{\rm h}32^{\rm m}15.470^{\rm s}$
 and Dec.=$+10^{\circ}59'29.01"$ (J2000). After background subtraction, there 
 were $29\pm5$ and $41\pm7$ net counts extracted from the arc-like feature, implying 
 the net counting rates of $(1.29\pm 0.26)\times10^{-3}$ cts/s and 
 $(1.65\pm 0.28)\times10^{-3}$ cts/s for the observations in 2005 December and 
 2006 May, respectively.  Within the $1\sigma$ errors of these count rates, no 
 variability is apparent on the basis of these two observations. 
 The response files were computed in the same manner as those for the pulsar spectra. 
 Each spectrum was dynamically binned so as to have at least 10 counts per bin. In order to 
 obtain a better statistic, we analyzed both spectra simultaneously. 

 We hypothesized that the nebular emission originates from the interaction of the 
 pulsar wind and the ISM. Synchrotron radiation from the ultra-relativistic 
 electrons is generally believed to be the emission mechanism of the pulsar 
 wind nebula, which is characterized by a power-law spectrum.  We tested this 
 hypothesis by fitting an absorbed power-law model to the nebular spectra. 
 We fixed the column density at the value $N_{H}=2.4\times 10^{21}$ cm$^{-2}$
 inferred from fitting the pulsar spectrum. The power-law model describes the 
 observed spectrum reasonably well ($\chi_{\nu}=0.88$ for 6 D.O.F.). 
 The best-fit power-law spectrum and residuals are shown in Fig.~7. 
 This model yields a photon index of $\Gamma=2.00^{+0.32}_{-0.30}$ 
 and a normalization at 1 keV of $3.63^{+0.78}_{-0.74}\times 10^{-6}$ photons
 keV$^{-1}$ cm$^{-2}$ s$^{-1}$. The unabsorbed flux 
 deduced for the best-fitted model parameters are $f_{X}=1.7\times 10^{-14}$ 
 erg s$^{-1}$ cm$^{-2}$ in the energy range of $0.5-10$ keV. The pulsar distance 
 of 361 pc implies a luminosity of $L_{X}=2.7\times 10^{29}$ erg s$^{-1}$. 

 We have checked the robustness of all the spectral parameters quoted here 
 by incorporating background spectra extracted from different source-free regions. We 
 found that within the $1\sigma$ errors the spectral parameters inferred 
 from independent fittings are all consistent with each other. 

\begin{figure}
\centerline{\psfig{figure=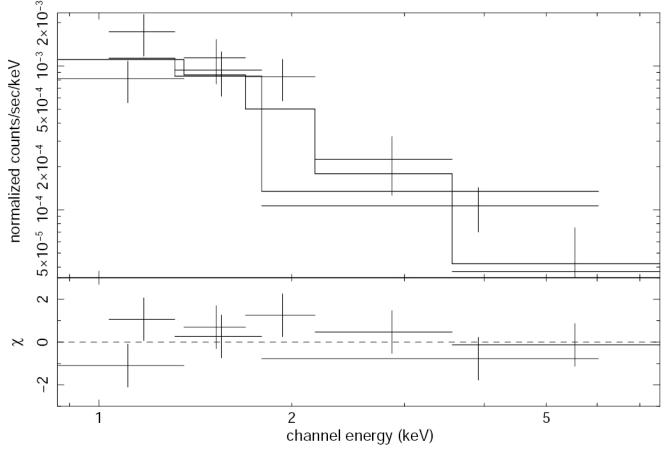,width=9cm,clip=}}
\caption{Energy spectra of the arc-like structure associated with 
\PSR\ as observed by the Chandra
ACIS-S3 detector on 04 Dec 2005 and 28 May 2006, and simultaneously
fitted to an absorbed power-law model
({\it upper panel}) and contribution to the $\chi^{2}$ fit statistic
({\it lower panel}).}
\end{figure}

\section{Discussion}
 
 Complementing the XMM-Newton observation, we have studied \PSR\ and its 
 surrounding medium with Chandra, which provides data with sub-arcsecond resolution. 
 So far, three distinct components have been resolved, namely the pulsar itself, the
 X-ray trail opposite to the pulsar's proper motion direction as well as the arc-like 
 nebula surrounding the pulsar. The flux contributions from these components 
 and the corresponding X-ray conversion efficiencies are summarized in Table 3. 

\begin{table}
\caption{Unabsorbed fluxes, luminosities, and the conversion efficiencies of \PSR\ 
and the nebular components in $0.5-10$ keV}
\begin{center}
\begin{tabular}{lccc}
\hline\hline
Component & $f_{X}$ & $L_{X}$$^{a}$ & $L_{X}/\dot{E}$ \\
           & ergs s$^{-1}$ cm$^{-2}$ & ergs s$^{-1}$ & \\
\hline
\\
\PSR\              & $2.5\times 10^{-13}$ & $3.9\times 10^{30}$  & $1.0\times 10^{-3}$  \\
\\
Trail-like feature $^{b}$ & $5.3\times 10^{-14}$ & $8.3\times 10^{29}$  & $2.1\times 10^{-4}$  \\
\\
Arc-like feature   & $1.7\times 10^{-14}$ & $2.7\times 10^{29}$  & $6.9\times 10^{-5}$   \\
\\
\hline\hline
 \end{tabular}
 \end{center}
$^{a}$ {\footnotesize Luminosities are calculated at the assumed pulsar distance of 361 pc}\\
$^{b}$ {\footnotesize Adopted from Becker et al. (2006)}
\end{table}

The morphology of the arc-like nebula and the orientation with respect to the 
pulsar's proper motion direction suggest a bow-shock nature. 
This is the first well-defined bow-shock morphology observed from an old pulsar. 

A bow-shock nebula as observed in \PSR\ can be produced by a pulsar in supersonic motion. 
In this case,  the termination shock radius $R_{s}$ is determined by the balance of the 
ram pressure between the relativistic pulsar wind particles and the ISM at the head of the shock 
(cf. Hui \& Becker 2007b):
\begin{equation}
R_{s}\simeq(\dot{E}/2\pi\rho_{ISM}v_{p}^{2}c)^{1/2}
\sim3\times10^{16}\dot{E}_{34}^{1/2}n^{-1/2}v_{p,100}^{-1} {\rm~cm}
\end{equation}
where $v_{p,100}$ is the velocity of the pulsar in units of 100 km s$^{-1}$; 
$\dot{E}_{34}$ is the spin-down luminosity of the pulsar in units of $10^{34}$
erg s$^{-1}$; and $n$ is the number density of the ISM in units of cm$^{-3}$. 
A density of $1$ cm$^{-3}$ implies a shock radius of $R_{s}\sim10^{16}$ cm. 
The stand-off angle $\theta_{s}$ between the pulsar and edge of the shock 
in the forward direction is related to $R_{s}$ by $\theta_{s}=R_{s}\cos i/D$, where 
$D$ is the distance to the pulsar and 
$i$ is the inclination angle between the proper motion and the plane of the sky. Taking 
the FWHM resulted from Gaussian fitting to the radial profile (i.e., 1.13 arcsec) as the 
lower limit of $\theta_{s}$,  $i$ is then constrained to be $\lesssim50^{\circ}$. 

At a distance of 361 pc, the proper motion velocity and the size of the arc-like nebula imply 
the pulsar took $\sim 150$ years to pass through the sky region of the nebula. 
We estimated the magnetic field in this shocked region by equating the passage time to the 
synchrotron cooling timescale $\tau_{\rm syn}=6\pi m_{e}c/\gamma\sigma_{T}B^{2}. 
\simeq 10^{5}\left(h\nu/{\rm keV}\right)^{-0.5}B_{\mu G}^{-3/2}$ yr, 
where $\gamma$ is the Lorentz factor of the wind, taken to be
$10^{6}$ (cf.,~Cheng et al.~2004), $\sigma_{T}$ is the Thompson cross
section, and $B_{\mu G}$ is the magnetic field in the shocked region in units of micro gauss. 
This implies a magnetic field of $\sim 75$ $\mu G$ for the region near to the pulsar. 
For comparison, the typical field 
strength in the ISM is $\sim 2-6$ $\mu G$ (cf.,~Beck et al.~2003, and references therein). 
If the magnetic field in the neighborhood of \PSR\ is comparable 
with this estimate, the inferred compression factor is $\sim 13-38$ in the 
bow-shock region, which is rather high in comparison with other pulsar wind 
nebulae (e.g., Hui \& Becker 2006, 2007b). Polarization measurements of the 
sky region around \PSR\ in the radio band can help to better constrain the 
magnetic field of this system and hence the compression factor. 

In the context of the standard shock theory (Chevalier 2000), the X-ray
luminosity and spectral index depend on the inequality between the characteristic
observed frequency $\nu_{X}^{\rm obs}$ and the electron synchrotron cooling
frequency:
$\nu_{\rm c}=18\pi em_{e}c/\sigma_{T}^{2}\tau^{2}_{\rm syn}B^{3}$
(cf.,~Chevalier 2000 and references therein), which is estimated to be 
$\nu_{\rm c}=1.8\times10^{17}$ Hz. Since, in general $\nu_{X}^{\rm obs}>\nu_{c}$, 
this suggests the X-ray emission is in a fast cooling regime. 

We further probed the energy distribution of the synchrotron radiating 
electrons in the shock. The emitting electrons are distributed as 
$N\left(\gamma\right)\propto\gamma^{-p}$. In a fast cooling regime, the 
theoretical luminosity per unit frequency is given by (cf.,~Cheng et 
al.~2004):

\begin{equation}
L^{\rm th}_{\nu,p}=\frac{1}{2}\left(\frac{p-2}{p-1}\right)^{p-1}
\left(\frac{6e^{2}}{4\pi^{2}m_{e}c^{3}}\right)^{\frac{p-2}{4}}
\epsilon_{e}^{p-1}\epsilon_{B}^{\frac{p-2}{4}}\gamma^{p-2}
R_{s}^{-\frac{p-2}{2}}\dot{E}^{\frac{p+2}{4}}\nu^{-\frac{p}{2}}.
\end{equation}
Assuming the energy equipartition between the electron and proton, we took the
fractional energy density of electrons $\epsilon_{e}$ to be $\sim 0.5$ and the
fractional energy density of the magnetic field $\epsilon_{B}$ to be $\sim 0.01$.
We then integrated Eq.~2 over the frequency: 
$L^{\rm th}_{p}=\int_{\nu_{min}}^{\nu_{max}} L^{\rm th}_{\nu,p}d\nu$,
with $h\nu_{min}=$0.5~keV and $h\nu_{max}=$10~keV. Equating $L^{\rm th}_{p}$ and 
the observed luminosity of the arc-like nebula results in an index of $p=2.35$ for the 
electron spectrum. This estimate lies within the acceptable range of the index $p$ as suggested 
in a standard shock model, i.e.,~$p\sim2-3$ (cf.,~Cheng et al.~2004 and references therein). 
For a fast cooling scenario, the photon index is related to $p$ as $\Gamma^{th}=(p+2)/2$. 
The value $p=2.35$ implies a photon index of $\Gamma^{th}=2.18$. This theoretical value lies 
within a $1\sigma$ uncertainty of the observed value. 

The arc-shaped morphology of the nebula observed in \PSR\ is similar to that of Geminga 
(Caraveo et al.~2003). However, there is a main difference between these two cases. 
While the arc-like nebula of Geminga is rather symmetric with respect to the direction 
of proper motion, asymmetry is indicated in the case of \PSR\ (see Fig.~1b). Its 
asymmetric shape might be a result of the anisotropic pulsar wind and/or inhomogeneities 
in the surrounding ISM (see Hui \& Becker 2006). 

For the pulsar \PSR,  we found that the spectral properties inferred from our analysis are in good 
agreement with the results obtained by XMM-Newton, which did not allow the 
pulsar emission to be separated from the compact nebular component. The consistency is 
expected even in the presence of the diffuse compact nebulae as it contributes only  
$\sim 7\%$ to the observed energy flux (see Table 3). Thus, the non-thermal 
emission scenario of \PSR\ is confirmed.  Hui \& Becker (2007a) had argued that 
it is possible to sustain particle acceleration regions in \PSR's outer-magnetosphere 
(so-called outer-gap) if the inclination of the magnetic axis with respect to the rotational 
axis is taken into account.  This inference is supported by the ability of
the outer-gap model in reproducing the observed X-ray 
pulse profile and its phase shift relative to the radio pulse (see Fig.~16 Becker et
al.~2006). However, this model has difficulty in explaining the observed spectral 
properties of \PSR.
 
In the outer-gap model, the non-thermal X-rays result from the back-flowing charge particles  
from the outer gap (Cheng \& Zhang 1999). When the primary electrons/positrons leave the outer-gap, 
they will emit curvature photons that are subsequently converted into secondary pairs in the 
presence of the strong magnetic field. 
Synchrotron photons will then be emitted by these secondary electrons/positrons. If these photons are 
energetic enough, they will further be converted into pairs that again lose their energy via 
synchrotron radiation. Therefore, an electromagnetic cascade is developed. Based on this model, 
Cheng \& Zhang (1999) argued that the X-ray photon index resulting from such cascades  
should be $\leq 2$. This is obviously not 
in agreement with the observed photon index for \PSR, which is as steep as $2.91^{+0.16}_{-0.13}$. 
Observations of five other old pulsars: B1133+16 (Kargaltsev,Pavlov \& Garmire 2006); 
B0943+10 (Zhang, Sanwal \& Pavlov 2005); B0628-28 (Becker et al.~2005); 
B0823+26 (Becker et al.~2004); and J2043+2740 (Becker et al.~2004), 
also found the photon indices steeper than 2. 
This gives the outer-gap emission model a challenge. Re-examination of the model is thus required. 

After this work was submitted to A\&A for publication, we became aware
that a paper on the similar subject as presented here was submitted to
ApJ by Misanovic, Pavlov, \& Garmire (2007). The results reported by
these authors agree well with the results presented here. Misanovic
et al., though, report to have detected a thermal component in the
spectrum of \PSR. While their parameters inferred from the single
power-law fittings are consistent with ours, they suggested that a
model of power-law plus blackbody model provides a statistically
better description. This obviously contradicts with the conclusion
presented here and in Becker et al.~(2006). From a closer inspection
of their spectral fits we conclude that their apparent detection is
probably biased by a too small signal-to-noise ratio per spectral bin.
While we binned the Chandra spectrum so as to have at least 30 counts
per bin Misanovic et al.~(2007) grouped the data so as to have only
about 15 counts per spectral bin. For the spectral fitting of XMM-Newton
data, the parameters obtained by Misanovic et al.~(2007) are for 190 degree
of freedom whereas Becker et al.~(2006) had only 121 degrees of freedom
for their analysis. Becker et al.~(2006) thus used a much higher photon 
statistics per spectral bin. 
The higher signal-to-noise ratio per spectral bin yielded a more 
stringent discrimination among competing spectral models. For this reason, 
the spectral results presented here as well as in Becker et al.~(2006) 
provide more stringent constraints on the pulsar's X-ray emission 
properties than those in Misanovic et al.~(2007).

\begin{acknowledgements}
We thank the referee Patrizia Caraveo for providing many useful suggestions to 
improve the quality of the manuscript considerably. 
We would also thank Bernd Aschenbach for thoroughly reading the manuscript 
and K. S. Cheng for discussing the shock physics with us.
\end{acknowledgements}

\end{document}